\documentclass[pra,aps,eqsecnum,amsmath,amssymb,showkeys]{revtex4}
\usepackage[mathscr]{eucal}
\usepackage{xtheorem}
\newcommand{\xx}{\mathbf{x}}
\newcommand{\yy}{\mathbf{y}}
\newcommand{\pq}{\mathbf{p}}
\newcommand{\qq}{\mathbf{q}}
\newcommand{\rr}{\mathbf{r}}
\newcommand{\bro}{\boldsymbol{\rho}}
\newcommand{\bva}{\boldsymbol{\varrho}}
\newcommand{\bom}{\boldsymbol{\omega}}
\newcommand{\tbro}{\boldsymbol{\Tilde{\rho}}}
\newcommand{\pbi}{\openone}
\newcommand{\veps}{\varepsilon}
\newcommand{\hh}{\mathcal{H}}
\newcommand{\kk}{\mathcal{K}}

\newtheorem{plain}{Thm}{Theorem}[section]
{Lemma}
{Definition}
{Remark}

\begin{document}
\clearpage
\preprint{}

\title{Continuity and Stability of Partial Entropic Sums}
\author{Alexey E. Rastegin}
\affiliation{Department of Theoretical Physics, Irkutsk State University,
Gagarin Bv. 20, Irkutsk 664003, Russia}

\begin{abstract}
Extensions of Fannes' inequality with partial sums of the Tsallis
entropy are obtained for both the classical and quantum cases. The
definition of $k$-th partial sum under the prescribed order of
terms is given. Basic properties of introduced entropic measures
and some applications are discussed. The derived estimates provide
a complete characterization of the continuity and stability
properties in the refined scale. The results are also reformulated
in terms of Uhlmann's partial fidelities.
\end{abstract}


\maketitle

\section{Introduction}

The concept of entropy is much widely used in statistical
mechanics, quantum physics and information theory. In the last
decades, new direction of researches has been formed on the
joining point of these fields. The discipline is usually referred
to as ''quantum information'' \cite{chuang,petz08}. In all the
topics, it is necessary to quantify the informational contents of
a state by some functionals. For these purposes, various entropic
quantities have been utilized \cite{ohya,petz08}. The Shannon
entropy for probability distributions and the von Neumann entropy
for density operators are widely adopted measures. Many other
functionals like the Tsallis entropies \cite{havrda,tsallis}, the
R\'{e}nyi entropies \cite{renyi}, and the quasi-entropies
\cite{petz86} have been recognized as needed in specialized
topics. Quantum entropic measures are nontrivial mathematical
subjects, and active study of them has led to some deep insights
(see, e.g., historical remarks in \cite{carlen08}).

Any entropic measure must have good functional properties. Fannes
showed the continuity property for the von Neumann entropy
\cite{fannes}. Fannes' inequality gives an upper bound on a
potential change of the entropy when the quantum state is altered.
Note that the dimension of the state space is explicitly involved
in the bound. Using the classical Fano inequality, Fannes' bound
has been sharpened (see theorem 3.8 in \cite{petz08}). Recently,
its analogs for Tsallis entropies were obtained
\cite{yanagi,zhang}. Further, the stability property is very
important, mainly due to its direct relation to observability
\cite{abe02}. In spite of direct formal relation between the
Tsallis $\alpha$-entropy and the R\'{e}nyi $\alpha$-entropy, the
former is stable \cite{abe02,zhang}, while the latter is not
stable for $\alpha\not=1$ \cite{abe02,lesche}. There are other
important properties, for example, related to nonextensive
additivity. The quantum Tsallis entropy of order $\alpha>1$ is
subadditive (that has been conjectured in \cite{raggio} and later
proved in \cite{auden07}). Concerning the minimum relative-entropy
principle, an extension of triangle inequality to the nonextensive
case was developed \cite{dmb}. In the present paper, we focus an
attention on the continuity and stability properties.

Functional properties may be asked with respect to separate
components in expression for entropy. A study of partial sums of
some measure may be useful in various regards. By means of the Ky
Fan norms, many important results can be extended to the class of
unitarily invariant norms. For instance, all the metrics induced
by unitarily invariant norms are tantamount in posing the
cryptographic exponential indistinguishability \cite{rast10}. The
equivalence of pairs of density operators has been resolved in
terms of Uhlmann's partial fidelities \cite{uhlmann00}. In view of
very importance of both the continuity and stability, one of
possible development of the issue is to examine these properties
for partial entropic sums. The paper is organized as follows. In
Section \ref{clasps}, we deal with the classical case and review
auxiliary material. In Section \ref{quanps}, the case of density
operators with some applications is studied. In Sections
\ref{alptwo} and \ref{alpinf} the main result will be established
for two different ranges of parameter $\alpha$. In an explicit
form, the bounds on differences between $k$-th partial entropic
sums are dependent on integer $k$ only. In this respect, the
obtained inequalities differ from Fannes' inequality. We also show
that examined partial sums are stable in the sense inspired by
Lesche \cite{lesche}.

\section{Definitions for classical distributions}\label{clasps}

Let $\xx=(x_1,\ldots,x_m)$ be an element of real space
${\mathbb{R}}^m$. For $k=1,\ldots,m$, we define the function
\begin{equation}
\label{gaugf}
G_{(k)}(\xx):=\sum\nolimits_{i=1}^{k}|x_i|^{\downarrow}
\ ,
\end{equation}
where the arrows down indicate that absolute values are put in the
decreasing order. Note that $G_{(m)}(\xx)$ poses $l_1$-norm and
$G_{(1)}(\xx)$ poses $l_{\infty}$-norm on ${\mathbb{C}}^m$
\cite{bhatia}. In terms of these symmetric gauge functions, the Ky
Fan norms of operators are defined. Let $x\mapsto f(x)$ be a
function of real variable. For any
$f:[0;1]\rightarrow{\mathbb{R}}_{+}\equiv[0;+\infty)$, we
introduce the map $\xx\mapsto{G}_{(k)}[f(\xx)]$ by
\begin{equation}
G_{(k)}[f(\xx)]:=\sum\nolimits_{i=1}^{k}|f(x_i)|^{\downarrow} \ .
\end{equation}
Here a vector ${\xx}$ is assumed to be contained in the
probability simplex $\Delta_m$ of those vectors that
$G_{(m)}(\xx)=1$ and $x_i\geq0$ for all $i$. Let us put the
${\alpha}$-entropy function \cite{tsallis}
\begin{equation}
\label{qentf}
\eta_{\alpha}(x):=-x^{\alpha}\ln_{\alpha}(x)=\frac{x^{\alpha}-x}{1-{\alpha}} \ ,
\end{equation}
where the ${\alpha}$-logarithmic function
$\ln_{\alpha}(x)\equiv(x^{1-{\alpha}}-1)/(1-{\alpha})$ is defined
for ${\alpha}\geq0$, ${\alpha}\not=1$ and $x\geq0$. The
${\alpha}$-logarithmic function converges to $\ln{x}$ as
${\alpha}\to1$. In the following, we avoid the case $\alpha=0$ in
which we have $\eta_{\alpha}(0)\neq0$.

\begin{Rem}\label{econc}
In the range $x\in[0;1]$ the function $\eta_{\alpha}(x)$ is
concave, $\eta_{\alpha}(0)=\eta_{\alpha}(1)=0$. Its maximum in
this range is reached at the point
$x_0(\alpha)=\alpha^{1/(1-\alpha)}$. It can be shown that
$dx_0/d\alpha>0$ for $\alpha\in[0;+\infty)$, and $x_0(1)=1/e$,
$x_0(2)=1/2$, $x_0(+\infty)=1$.
\end{Rem}

\begin{Def}\label{hkp}
Let $\pq=(p_1,\ldots,p_m)$ be $m$-dimensional
probability vector. For $k=1,\ldots,m$, the $k$-th partial
entropic sum is defined by
\begin{equation}
\label{defclas}
H_{\alpha}^{(k)}(\pq):=G_{(k)}[\eta_{\alpha}(\pq)] \ .
\end{equation}
\end{Def}

In the case $k=m$, we obtain the entropy itself. This measure is
usually referred to as ''Tsallis entropy'', although it was first
studied by Havrda and Charv\'{a}t \cite{havrda}. The Tsallis
entropy has many applications in nonextensive statistical
mechanics. Unlike the Shannon entropy and the R\'{e}nyi entropy,
the Tsallis entropy is not additive \cite{auden07,raggio}. The partial sums
of Tsallis entropy enjoy a few obvious properties.
\begin{enumerate}
 \item[(1C)]{{\it Positivity:} $H_{\alpha}^{(k)}(\pq)\geq0$; $H_{\alpha}^{(k)}(\pq)=0$ only in corners of the simplex $\Delta_m$.}
 \item[(2C)]{{\it Non-decrease with respect to the order:} if $k<k'$ then $H_{\alpha}^{(k)}(\pq)\leq H_{\alpha}^{(k')}(\pq)$.}
 \item[(3C)]{{\it Symmetry:} if $\qq$ is obtained by permutation of $\pq$ then $H_{\alpha}^{(k)}(\qq)=H_{\alpha}^{(k)}(\pq)$.}
 \item[(4C)]{{\it Expansibility:} $H_{\alpha}^{(k)}(p_1,\ldots,p_m,0)=H_{\alpha}^{(k)}(p_1,\ldots,p_m)$.}
 \item[(5C)]{{\it Product monotonicity:} $H_{\alpha}^{(k)}(\pq)\leq H_{\alpha}^{(kn)}(\pq\times\qq)$, where $\pq\times\qq$
  denotes the joint probability distribution with elements $p_iq_j$ ($j=1,\ldots,n$).}
\end{enumerate}
These properties follow from Definition \ref{hkp} at once. Only the last
demands some comments. In effect, a stronger statement takes
place.

\begin{Lem}\label{monotl}
Let $\rr=\{r_{ij}:1\leq i\leq m,\>1\leq j\leq n\}$ be a
probability distribution. Then it holds that
\begin{equation}
H_{\alpha}^{(k)}(\pq)\leq H_{\alpha}^{(kn)}(\rr)
\ , \label{monclas}
\end{equation}
where $p_i=\sum_{j=1}^{n}r_{ij}$ are elements of the marginal
probability distribution.
\end{Lem}

{\it Proof.} Let ${\mathcal{A}}$ be $k$-subset of the set $\{1,\ldots,m\}$ such that
$H_{\alpha}^{(k)}(\pq)=\sum_{i\in{\mathcal{A}}}\eta_{\alpha}(p_i)$.
Then we have
\begin{align}
H_{\alpha}^{(k)}(\pq)&=\frac{1}{1-\alpha}\sum\nolimits_{i\in{\mathcal{A}}}\left(\sum\nolimits_{j=1}^{n}r_{ij}\right)^{\alpha}
-{\>}\frac{1}{1-\alpha}\sum\nolimits_{i\in{\mathcal{A}}}\sum\nolimits_{j=1}^{n}r_{ij} \nonumber\\
&\leq\frac{1}{1-\alpha}\sum_{i\in{\mathcal{A}}}\sum_{j=1}^{n}r_{ij}^{\alpha}
{\>}-{\>}\frac{1}{1-\alpha}\sum_{i\in{\mathcal{A}}}\sum_{j=1}^{n} r_{ij}=
\sum_{i\in{\mathcal{A}}}\sum_{j=1}^{n} \eta_{\alpha}(r_{ij})
\label{iaijan} \ ,
\end{align}
because
$\left(\sum_{j=1}^{n}r_{ij}\right)^{\alpha}\leq\sum_{j=1}^{n}r_{ij}^{\alpha}$
for $\alpha<1$, and
$\left(\sum_{j=1}^{n}r_{ij}\right)^{\alpha}\geq\sum_{j=1}^{n}r_{ij}^{\alpha}$
for $\alpha>1$. In (\ref{iaijan}), the last sum is taken over
those pairs $(i,j)$ that $i\in{\mathcal{A}}$ and
$1\leq{j}\leq{n}$. The number of such pairs is $kn$, whence this
sum is not greater than $H_{\alpha}^{(kn)}(\rr)$. $\square$

When $m$ and $\alpha$ are fixed, the maximum of the Tsallis entropy \cite{yanagi,zhang}
\begin{equation}
\max\left\{H_{\alpha}^{(m)}(\pq):{\>}\pq\in\Delta_{m}\right\}
=\ln_{\alpha}(m)
\label{maxts}
\end{equation}
is reached if and only if $p_i=1/m$ for all $i=1,\ldots,m$. It is
important that this optimal probability vector is independent of
$\alpha$. Conversely, for $k$-th partial entropic sum
$H_{\alpha}^{(k)}(\pq)$ the maximizing probability vector is
dependent on both $k$ and $\alpha$. To find explicitly the maximum
of $k$-th partial sum, we should maximize the function
$G_{(k)}[\eta_{\alpha}(\xx)]$ of vector ${\xx}\in{\mathbb{R}}^m$
under the conditions $x_i\geq0$ and $G_{(k)}(\xx)\leq1$. Without
equality $G_{(k)}(\xx)=1$, the task becomes difficult since the
function $\eta_{\alpha}(x)$ is not scale-invariant. Nevertheless,
we can give simple lower and upper bounds
\begin{equation}
\ln_{\alpha}(k)\leq\max\left\{H_{\alpha}^{(k)}(\pq):{\>}\pq\in\Delta_{m}\right\}\leq\ln_{\alpha}(k+1)
\label{maxpts}
\end{equation}
with high accuracy for sufficiently large $k$. More precisely, the
relative error of estimate by these bounds does not exceed
$(k^{\alpha}\ln_{\alpha}(k+1))^{-1}$. When $p_i=1/k$ for
$1\leq{i}\leq{k}$ and $p_i=0$ for $k+1\leq{i}\leq{m}$, $k$-th
partial sum is $\ln_{\alpha}(k)$, whence the lower bound follows.
Further, for each probability vector $\pq$ with $G_{(k)}(\pq)<1$
we take another one $\pq'$ such that $p_i^{\>\prime}=p_i$ for
$i=1,\ldots,k$, and $p_{k+1}^{\>\prime}=1-G_{(k)}(\pq)$. By
construction, we now have $G_{(k)}(\pq')=1$ and
\begin{equation}
H_{\alpha}^{(k)}(\pq)\leq H_{\alpha}^{(k+1)}(\pq')\leq\ln_{\alpha}(k+1)
\ , \label{pqchain}
\end{equation}
whence the upper bound follows. For $k=1$, the lower bound in
(\ref{maxpts}) becomes trivial. Here we will use the exact
expression for the maximum, namely
$\eta_{\alpha}(\alpha^{1/(1-\alpha)})$.

\begin{Rem}\label{a11a}
Since $\alpha^{1/(1-\alpha)}\to{e}^{-1}$ as $\alpha\to1$, the
maximum of second partial sum $H_{1}^{(2)}$ is equal to
$2\eta_{1}(1/e)=2/e\approx0.74$. This is larger than the maximal
value $\ln{2}\approx0.69$ of the binary Shannon entropy. So the
left-hand side of (\ref{maxpts}) is lower bound too. For partial
sums of the Shannon entropy, the relative error of estimate by
(\ref{maxpts}) is almost $0.1$ for $k=5$ and strictly less than
$0.1$ for $k>5$.
\end{Rem}

What is a reason for separate analysis of a partial sums
stability? Consider the two-dimensional probability vectors
$\pq=\bigl((1-\veps)/2,(1+\veps)/2\bigr)$ with strictly positive
$\veps\ll1$ and $\qq=\bigl(1/2,1/2\bigr)$. The $\eta_1(x)=-x\ln x$
is maximal for $x=1/e$, whence
$\eta_1\bigl((1-\veps)/2\bigr)>\eta_1\bigl((1+\veps)/2\bigr)$.
Putting
$\Delta^{(k)}=\bigl|H_{1}^{(k)}(\pq)-H_{1}^{(k)}(\qq)\bigr|$, we
write
\begin{align}
 & \Delta^{(1)}=\eta_1\bigl((1-\veps)/2\bigr)-\eta_1\bigl(1/2\bigr)
 =\bigl(-(1-\veps)\ln(1-\veps)-\veps\ln2\bigr)\big/2
\ , \label{dif1}\\
 & \Delta^{(2)}=H_{1}^{(2)}(\qq)-H_{1}^{(2)}(\pq)
=\bigl((1+\veps)\ln(1+\veps)+(1-\veps)\ln(1-\veps)\bigr)\big/2
\ . \label{dif2}
\end{align}
The ratio of $\Delta^{(1)}$ to $\Delta^{(2)}$ is equal to
$(1-\ln2){\,}\veps^{-1}+O(1)$ and not bounded as $\veps\to+0$. Thus, the
difference between partial entropic sums can be arbitrarily large in
comparison with the difference between the total entropies. We
have seen this effect in a vicinity of the uniform distribution.
So the continuity and stability of some entropic functional
themselves do not imply the same for corresponding partial sums.
It turns out, however, that these properties still hold for
partial sums of Tsallis' entropy. So we will examine differences
between $k$-th partial sums of Tsallis entropy of two
$m$-dimensional probability vectors $\pq$ and $\qq$. It is natural
to estimate this difference in terms of partitioned classical
distances $G_{(k)}(\pq-\qq)$. These distances and their quantum
extensions via Ky Fan's norms were introduced in \cite{rast091}.
As it is shown below, both the classical and quantum cases can
similarly be treated.

\section{The case of density operators}\label{quanps}

Let ${\hh}$ be $d$-dimensional Hilbert space, and let
${\mathsf{X}}$ be linear operator on ${\hh}$. By
${\rm{spec}}({\mathsf{X}})\equiv\{\lambda_i({\mathsf{X}})\}$ we
denote the set of its eigenvalues. For any ${\mathsf{X}}$, the
operator ${\mathsf{X}}^{*}{\mathsf{X}}$ is positive, i.e.
$\langle\psi|{\mathsf{X}}^{*}{\mathsf{X}}|\psi\rangle\geq0$ for
all $|\psi\rangle\in{\hh}$. The operator $|{\mathsf{X}}|$
is defined as a unique positive square root of
${\mathsf{X}}^{*}{\mathsf{X}}$. The eigenvalues of
$|{\mathsf{X}}|$ counted with multiplicities are the singular
values $\sigma_j({\mathsf{X}})$ of ${\mathsf{X}}$ \cite{bhatia}.
For $k=1,\ldots,d$, the Ky Fan $k$-norm is defined as
\cite{bhatia}
\begin{equation}
\|{\mathsf{X}}\|_{(k)}:=G_{(k)}(\sigma({\mathsf{X}}))
\equiv\sum\nolimits_{j=1}^{k} \sigma_j^{\downarrow}({\mathsf{X}})
\ . \label{fannorm}
\end{equation}
For given $f(x)$, the function of normal operator
${\mathsf{X}}=\sum_{i=1}^d \lambda_i|e_i\rangle\langle{e_i}|$ is
introduced by
\begin{equation}
f({\mathsf{X}})=\sum\nolimits_{i=1}^d f(\lambda_i)|e_i\rangle\langle{e_i}|
\ , \label{fnopdf}
\end{equation}
where the eigenvectors $|e_i\rangle$ are orthonormal. We also have
$\|{\mathsf{X}}\|_{(k)}=G_{(k)}(\lambda({\mathsf{X}}))$ for each
normal operator ${\mathsf{X}}$.

\begin{Def}
Let $\bro$ be density operator on
${\hh}$. For $k=1,\ldots,d$, the $k$-th partial entropic
sum of quantum state $\bro$ is defined by
\begin{equation}
\label{defquan}
S_{\alpha}^{(k)}({\bro}):=\|\eta_{\alpha}({\bro})\|_{(k)} \ .
\end{equation}
\end{Def}

\begin{Rem}
The eigenvalues $\{p_i\}$ of density operator
${\bro}$ are naturally treated as probabilities due to their
positivity and the normalization ${\rm{tr}}({\bro})=1$. Because
the function $\eta_{\alpha}(x)$ is nonnegative for $x\in[0;1]$,
the above definition and (\ref{fnopdf}) lead to
\begin{equation}
S_{\alpha}^{(k)}({\bro})=H_{\alpha}^{(k)}({\pq})
\ . \label{quaclas}
\end{equation}
\end{Rem}

In the case $k=d$, we have the quantum Tsallis entropy
\cite{yanagi,zhang}. The quantum partial entropic sums enjoy
similar properties to classical ones.
\begin{enumerate}
 \item[(1Q)]{{\it Positivity:} $S_{\alpha}^{(k)}({\bro})\geq0$; $S_{\alpha}^{(k)}({\bro})=0$ if and only if $\bro$ is projector.}
 \item[(2Q)]{{\it Non-decrease with respect to the order:} if $k<k'$ then $S_{\alpha}^{(k)}({\bro})\leq S_{\alpha}^{(k')}({\bro})$.}
 \item[(3Q)]{{\it Symmetry:} if ${\rm{spec}}({\bva})={\rm{spec}}({\bro})$ then $S_{\alpha}^{(k)}({\bva})=S_{\alpha}^{(k)}({\bro})$.}
 \item[(4Q)]{{\it Expansibility:} if the Hilbert space ${\hh}$ is extended to ${\hh}\oplus{\kk}$ then $S_{\alpha}^{(k)}(\bro)$ is not changed for all density operators $\bro$ on ${\hh}$.}
 \item[(5Q)]{{\it Product monotonicity:} $S_{\alpha}^{(k)}({\bro})\leq S_{\alpha}^{(kN)}({\bro}\otimes{\bom})$, where $\bom$ denotes some density operator on $N$-dimensional Hilbert space.}
\end{enumerate}
The entry (3Q) contains the unitary invariance as a particular case.
Like (5C), the property (5Q) can be extended to density operators
of a specific kind. Suppose a composite 'AB' of systems 'A' and 'B'
is described by density operator $\tbro$ on the tensor product
${\hh}_A\otimes{\hh}_B$. The reduced density
operators for systems 'A' and 'B' are given by
\begin{equation}
\bro_A={\rm{tr}}_B({\tbro}) \ ,\quad \bro_B={\rm{tr}}_A({\tbro})
\ , \label{rhotil}
\end{equation}
where the partial traces are taken over ${\hh}_B$ and
${\hh}_A$ respectively \cite{chuang,petz08}. In the two
spectral decompositions
\begin{equation}
\bro_A=\sum\nolimits_{i=1}^d a_i|i\rangle\langle{i}|
\ ,\quad \bro_B=\sum\nolimits_{\mu=1}^N b_{\mu}|\mu\rangle\langle\mu|
 , \label{spaspb}
\end{equation}
the $|i\rangle$'s form an orthonormal basis in ${\hh}_A$,
the $|\mu\rangle$'s form an orthonormal basis in
${\hh}_B$, ${\rm{spec}}({\bro_A})=\{a_i\}$ and
${\rm{spec}}({\bro_B})=\{b_{\mu}\}$. Then the vectors
$|i\mu\rangle\equiv|i\rangle\otimes|\mu\rangle$ form orthonormal
basis in the product space
${\hh}_A\otimes{\hh}_B$.

\begin{Thm}\label{tens}
If there hold: (i) the operators ${\tbro}$ and $\bro_A\otimes\bro_B$ are commuting, and (ii)
$a_ib_{\mu}={a}_jb_{\nu}$ only if ${\>}i=j$ and $\mu=\nu$, then
\begin{equation}
S_{\alpha}^{(k)}({\bro}_A)\leq S_{\alpha}^{(kN)}({\tbro})
\ , \quad S_{\alpha}^{(k)}({\bro}_B)\leq S_{\alpha}^{(kd)}({\tbro})
\ . \label{postl}
\end{equation}
\end{Thm}

{\it Proof.}
In the fixed basis $\{|i\mu\rangle\}$, the operator ${\tbro}$ can be represented as
\begin{equation}
{\tbro}=\sum\nolimits_{i\mu}\sum\nolimits_{j\nu} c(i\mu|j\nu)\>|i\mu\rangle\langle{j}\nu|
\ ,  \label{brrepr}
\end{equation}
where coefficients
$c(i\mu|j\nu)=\langle{i}\mu|{\tbro}|j\nu\rangle$. Further, the
product $\bro_A\otimes\bro_B$ is given by
\begin{equation}
\bro_A\otimes\bro_B=\sum\nolimits_{i\mu} a_ib_{\mu}|i\mu\rangle\langle{i}\mu|
\ ,  \label{abrepr}
\end{equation}
that is related to diagonal matrix. Using this fact, we simply obtain
\begin{align}
{\tbro}{\>}(\bro_A\otimes\bro_B)&=\sum\nolimits_{i\mu}\sum\nolimits_{j\nu}
c(i\mu|j\nu)\>a_jb_{\nu}|i\mu\rangle\langle{j}\nu|
\ , \nonumber\\
(\bro_A\otimes\bro_B){\>}{\tbro}&=\sum\nolimits_{i\mu}\sum\nolimits_{j\nu}
a_ib_{\mu}c(i\mu|j\nu)\>|i\mu\rangle\langle{j}\nu|
\ . \nonumber
\end{align}
The commutator of operators ${\tbro}$ and $\bro_A\otimes\bro_B$ has zero matrix elements, whence
\begin{equation}
\left(a_jb_{\nu}-a_ib_{\mu}\right)c(i\mu|j\nu)=0
\label{ajbn}
\end{equation}
for all values of labels. Under the precondition (ii) of theorem,
it follows from (\ref{ajbn}) that off-diagonal elements
$c(i\mu|j\nu)$ are all zero. Simplifying the notation to
$c_{i\mu}\equiv{c}(i\mu|i\mu)$, both the sets $\{a_i\}$ and
$\{b_{\mu}\}$ are marginal distributions of $\{c_{i\mu}\}$ due to
(\ref{rhotil}). Then the statement of Lemma \ref{monotl} completes the
proof. $\square$

\begin{Rem}
Let pair of qubits 'A' and 'B' be in the entangled pure state
$|\Psi(\theta)\rangle=\cos\theta|00\rangle+\sin\theta|11\rangle$.
Then ${\tbro}=|\Psi(\theta)\rangle\langle\Psi(\theta)|$ is
projector, and so its partial entropic sums are all zero. We also
have $\bro_A=a_0|0\rangle\langle0|+a_1|1\rangle\langle1|$ and
$\bro_B=b_0|0\rangle\langle0|+b_1|1\rangle\langle1|$, where
$a_0=b_0=(\cos\theta)^2$ and $a_1=b_1=(\sin\theta)^2$. Except for
$\cos2\theta=0$ and $\sin2\theta=0$, the operators ${\tbro}$ and
$\bro_A\otimes\bro_B$ are noncommuting. In this case, the
eigenvalues of both $\bro_A$ and $\bro_B$ are strictly positive,
whence $S_{\alpha}^{(k)}({\bro}_A)>0$ and
$S_{\alpha}^{(k)}({\bro}_B)>0$.
\end{Rem}

\begin{Rem}
When $\cos2\theta=0$, the operators ${\tbro}$
and $\bro_A\otimes\bro_B$ are commuting, but the inequalities
(\ref{postl}) are still not satisfied. Indeed, we have
$a_0=b_0=a_1=b_1=1/2$, $S_{\alpha}^{(k)}({\bro}_A)>0$, and
$S_{\alpha}^{(k)}({\bro}_B)>0$, while
$S_{\alpha}^{(k')}({\tbro})=0$ for all $k'$ as before. Here the
precondition (ii) of Theorem \ref{tens} is clearly violated. This
example shows an insufficiency of the precondition (i) itself. The
precondition (ii) is useful since only spectra of reduced
operators are involved. These spectra may be known from the
specification.
\end{Rem}

Another application of (\ref{monclas}) is related to quantum
measurements. A generalized measurement is described by ''positive
operator-valued measure'' (POVM). Recall that POVM
$\{{\mathsf{M}}_j\}$ is a set of positive operators
${\mathsf{M}}_j$ satisfying \cite{chuang,petz08}
\begin{equation}
\sum\nolimits_{j=1}^{N} {\mathsf{M}}_j={\pbi}
\ , \label{povmdef}
\end{equation}
where ${\pbi}$ is the identity in ${\hh}$. Following
\cite{davies}, we consider operators of rank one, i.e.
${\mathsf{M}}_j=|w_j\rangle\langle{w_j}|$. Let
$\{q_i,|\psi_i\rangle\}$ be one of those ensembles that generates
given density operator $\bva$. Generally, the vectors
$|w_j\rangle$ and $|\psi_i\rangle$ are neither normalized nor
orthogonal. Let $P(\psi_i;w_j)$ denote the joint probability that
the state $|\psi_i\rangle$ has been input and
$j$-th outcome has been obtained. For normalized $|w_j\rangle$, the term
$|\langle{w_j}|\psi_i\rangle|^2$ is the probability that
$|\psi_i\rangle$ passes the ''yes/no'' test of being the state
$|w_j\rangle$. Hence we obtain
\begin{align}
 & P(\psi_i;w_j)=q_i{\>}|\langle{w_j}|\psi_i\rangle|^2
\ , \label{condver}\\
 & S_{\alpha}^{(k)}({\bva})\leq H_{\alpha}^{(kN)}\big(P(\psi_i;w_j)\big)
\ , \label{meapp}
\end{align}
by using (\ref{monclas}) with $r_{ij}=P(\psi_i;w_j)$ and
$j=1,\ldots,N$. The above ''yes/no'' tests with pure states are
easily realized in physical experiments. In many tasks of quantum
information processing, a support and likewise spectrum sketch of
unknown mixed state are certain {\it a priori}. (Recall that
support of an operator is the subspace orthogonal to its kernel
\cite{chuang}.) So the relation (\ref{meapp}) may be applied in
design of feasible schemes to estinate the entropic sums in
practical specifications of such a kind. Here we consider a given
support as the actual Hilbert space in which the decomposition
(\ref{povmdef}) is taken. In principle, this issue might be a
subject of separate research.

\section{Inequalities for the case $\alpha\in(0;2]$}\label{alptwo}

In this section we establish a desired upper bound on difference
between two $k$-th entropic sums, when $\alpha\in(0;2]$. In
principle, the question can be recast as follows. For any two
$k$-dimensional real vectors $\xx$ and $\yy$, we define the
function
\begin{equation}
\label{fank}
F({\xx},{\yy}):=
\sum\nolimits_{i=1}^{k}\eta_{\alpha}(x_i)-
\sum\nolimits_{i=1}^{k}\eta_{\alpha}(y_i) \ .
\end{equation}
To show the continuity and stability properties for partial
entropic sums, we should obtain an upper bound on modulus of this
function in the domain
${\mathfrak{D}}_{\epsilon}\subset{\mathbb{R}}^k\times{\mathbb{R}}^k$
specified by
\begin{equation}
{\mathfrak{D}}_{\epsilon}:=\left\{({\xx},{\yy}):
x_i\geq0, \ y_i\geq0, \ G_{(k)}({\xx})\leq1, \ G_{(k)}({\yy})\leq1,
\ G_{(k)}({\xx}-{\yy})\leq\epsilon\right\}
\>. \label{domfank}
\end{equation}
Direct maximization of $|F({\xx},{\yy})|$ in
${\mathfrak{D}}_{\epsilon}$ seems to be difficult. Even if the
optimization is made for $({\xx},{\yy})\in\Delta_k\times\Delta_k$,
when both the $x_i$'s and the $y_i$'s are summarized to 1, the
task is very complicated. For instance, the function is neither
convex nor concave. Any extensions of admissible domain hamper a
solution. In the case $({\xx},{\yy})\in\Delta_k\times\Delta_k$,
the two complementary approaches are known.

The first method of estimating with restriction $\alpha\in[0;2]$
is presented in \cite{yanagi}. This is generalization of that
given for the von Neumann entropy in \cite{ohya} (see proposition
1.8 therein). By relevant modifications, we obtain the following
result.

\begin{Lem}\label{fxy}
If ${\>}\epsilon\leq\alpha^{1/(1-\alpha)}$ and
$\alpha\in(0;2]$ then in the domain $\mathfrak{D}_{\epsilon}$ it
holds that
\begin{equation}
|F({\xx},{\yy})|\leq\epsilon^{\alpha}\ln_{\alpha}(k+1)+\eta_{\alpha}(\epsilon)
\ . \label{kpart}
\end{equation}
\end{Lem}

{\it Proof.} If $x,y\in[0;1]$ and $|x-y|=t<1$, then the maximum of
$|\eta_{\alpha}(x)-\eta_{\alpha}(y)|$ is either $\eta_{\alpha}(t)$
or $\eta_{\alpha}(1-t)$. When $\alpha\in(0;2]$, the term
$h(t)\equiv\eta_{\alpha}(t)-\eta_{\alpha}(1-t)$ is positive for
$t\in[0;1/2]$, since $h(0)=h(1/2)=0$ and
$h^{\prime\prime}(t)\leq0$ for $t\leq1-t$. So, we have
\begin{equation}
|\eta_{\alpha}(x)-\eta_{\alpha}(y)|\leq\eta_{\alpha}(t)
\ . \label{contt}
\end{equation}
When $t\leq\alpha^{1/(1-\alpha)}$, by monotonicity this bound
holds for those points that $|x-y|\leq{t}$. Due to (\ref{contt}),
the quantity $|F({\xx},{\yy})|$ does not exceed
\begin{align}
 & \sum\nolimits_{i=1}^{k}|\eta_{\alpha}(x_i)-\eta_{\alpha}(y_i)|
\leq\sum\nolimits_{i=1}^{k}\eta_{\alpha}(\epsilon_i)=-\epsilon^{\alpha}\sum\nolimits_{i=1}^{k}
\frac{\epsilon_i^{\alpha}}{\epsilon^{\alpha}}\ln_{\alpha}\!\left(\frac{\epsilon_i}{\epsilon}\>\epsilon\right)
\nonumber\\
 & =-\epsilon^{\alpha}\sum_{i=1}^{k}\left\{ \left(\frac{\epsilon_i}{\epsilon}\right)^{\alpha}
 \ln_{\alpha}\!\left(\frac{\epsilon_i}{\epsilon}\right)
 +\frac{\epsilon_i}{\epsilon}\ln_{\alpha}(\epsilon) \right\}
 = \epsilon^{\alpha}\sum_{i=1}^{k}\eta_{\alpha}\!\left(\frac{\epsilon_i}{\epsilon}\right)
 + \eta_{\alpha}(\epsilon)\sum_{i=1}^{k}\frac{\epsilon_i}{\epsilon}
\label{fankmak}  \ ,
\end{align}
where we put $\epsilon_i\equiv|x_i-y_i|$ and use the identity
$\ln_{\alpha}(xy)=\ln_{\alpha}(x)+x^{1-\alpha}\ln_{\alpha}(y)$.
The last relations hold under the conditions $\alpha\in(0;2]$ and
$\sum_{i=1}^k\epsilon_i\leq\epsilon\leq\alpha^{1/(1-\alpha)}$.
Using the upper bound from (\ref{maxpts}) and
$\sum_{i=1}^k(\epsilon_i/\epsilon)\leq1$, we finally obtain
(\ref{kpart}). $\square$

\begin{Rem}
When $G_{(k)}(\xx-\yy)=\epsilon$, i.e.
$\sum_{i=1}^k(\epsilon_i/\epsilon)=1$, the set
$\{\epsilon_i/\epsilon\}$ is a probability distribution supported
on $k$ points. Then the multiplier of $\epsilon^{\alpha}$ in the
right-hand side of (\ref{fankmak}) is the Tsallis entropy itself,
whence
\begin{equation}
\sum\nolimits_{i=1}^{k}\eta_{\alpha}(\epsilon_i/\epsilon)\leq\ln_{\alpha}(k)\ ,
\quad |F({\xx},{\yy})|\leq\epsilon^{\alpha}\ln_{\alpha}(k)+\eta_{\alpha}(\epsilon)
\ . \label{kpoln}
\end{equation}
In fact, just the last was given in \cite{yanagi}. But we must
allow $G_{(k)}(\xx-\yy)<\epsilon$ for partial sums. Really, in the
three quantities $G_{(k)}[\eta_{\alpha}(\pq)]$,
$G_{(k)}[\eta_{\alpha}(\qq)]$, and $G_{(k)}(\pq-\qq)$ the index of
summation can run over three distinct $k$-subsets of the set
$\{1,\ldots,m\}$.
\end{Rem}

\begin{Rem}
If $\alpha>2$ then we have
$h^{\prime\prime}(t)>0$ and $h(t)<0$ for $0<t<1/2$. It is for this
reason that the case $\alpha>2$ cannot be analyzed within the
considered method.
\end{Rem}

Although the calculated bound is not sharp, it allows to establish
the continuity of partial sums. We pose the statement just for the
quantum case, because the case of classical distributions is
simultaneously treated. This is possible due to (\ref{quaclas})
and the following helpful result.

\begin{Lem}\label{gkq}
Let function $x\mapsto f(x)$ be positive-valued. For all ${\>}\pq,\qq\in{\mathbb{R}}^m$, there holds
\begin{equation}
\left|G_{(k)}[f(\pq)]-G_{(k)}[f(\qq)]+Q\right|\leq
\left|\sum\nolimits_{i\in{\mathcal{C}}}\bigl(f(p_i)-f(q_i)\bigr)+Q\right|
\ , \label{inelem0}
\end{equation}
where the $Q$ is independent of these sums, the ${\mathcal{C}}$ is a $k$-subset of the set $\{1,\ldots,m\}$.
\end{Lem}

{\it Proof.}
First, we introduce two $k$-subsets ${\mathcal{A}}$, ${\mathcal{B}}$ of the set $\{1,\ldots,m\}$ such that
\begin{equation}
G_{(k)}[f(\pq)]=\sum\nolimits_{i\in{\mathcal{A}}}f(p_i) \ ,
\qquad G_{(k)}[f(\qq)]=\sum\nolimits_{j\in{\mathcal{B}}}f(q_j)
\ . \label{inelem1}
\end{equation}
Putting intersection
${\mathcal{I}}={\mathcal{A}}{\cap}{\mathcal{B}}$, the term
$\left(G_{(k)}[f(\pq)]-G_{(k)}[f(\qq)]+Q\right)$ is recast as
\begin{equation}
\sum\nolimits_{i\in{\mathcal{I}}}\bigl(f(p_i)-f(q_i)\bigr)+
\sum\nolimits_{i\in({\mathcal{A}}\setminus{\mathcal{I}})}f(p_i)-
\sum\nolimits_{j\in({\mathcal{B}}\setminus{\mathcal{I}})}f(q_j)+Q
\ . \label{inelem2}
\end{equation}
The differences $({\mathcal{A}}\setminus{\mathcal{I}})$ and
$({\mathcal{B}}\setminus{\mathcal{I}})$ are sets of equal
cardinality. Without loss of generality, the value (\ref{inelem2})
can be assumed to be positive. Due to the definition of
${\mathcal{B}}$, we have $f(q_j)\geq f(q_i)$ for all
$j\in{\mathcal{B}}$ and $i\not\in{\mathcal{B}}$, including
$j\in({\mathcal{B}}\setminus{\mathcal{I}})$ and
$i\in({\mathcal{A}}\setminus{\mathcal{I}})$. Hence
$$
-\sum\nolimits_{j\in({\mathcal{B}}\setminus{\mathcal{I}})}f(q_j)
\leq-\sum\nolimits_{i\in({\mathcal{A}}\setminus{\mathcal{I}})}f(q_i) \ ,
$$
and the claimed statement is obtained by replacing the former sum
with the latter. Then we take ${\mathcal{C}}={\mathcal{A}}$ in the
inequality (\ref{inelem0}). $\square$

We are now able to prove the main result of this section. For two
density operators $\bro$ and $\bva$ on $d$-dimensional Hilbert
space ${\hh}$, we denote
$p_i=\lambda_i^{\downarrow}({\bro})$ and
$q_j=\lambda_j^{\downarrow}({\bva})$ with $i,j\in\{1,...,d\}$. In
this notation, we have
\begin{equation}
G_{(k)}[\pq-\qq]=G_{(k)}[\lambda^{\downarrow}({\bro})-\lambda^{\downarrow}({\bva})]\leq
G_{(k)}[\lambda({\bro}-{\bva})]=\|{\bro}-{\bva}\|_{(k)}
\ , \label{kfcon}
\end{equation}
The middle inequality is a particular case of theorem III.4.4 in
\cite{bhatia}. By Lemma \ref{gkq},
\begin{equation}
\left|S_{\alpha}^{(k)}({\bro})-S_{\alpha}^{(k)}({\bva})\right|=
\left|G_{(k)}[\eta_{\alpha}(\pq)]-G_{(k)}[\eta_{\alpha}(\qq)]\right|\leq
\biggl|\sum_{i\in{\mathcal{C}}}\bigl(\eta_{\alpha}(p_i)-\eta_{\alpha}(q_i)\bigr)\biggr|
\label{lemap}
\end{equation}
for some $k$-subset ${\mathcal{C}}$ of the set $\{1,...,d\}$. Using $\epsilon:=\sum_{i\in{\mathcal{C}}}|p_i-q_i|$
and $\varepsilon:=\|{\bro}-{\bva}\|_{(k)}$, the right-hand side of (\ref{lemap}) does not exceed
\begin{equation}
\epsilon^{\alpha}\ln_{\alpha}(k+1)+\eta_{\alpha}(\epsilon)\leq\varepsilon^{\alpha}\ln_{\alpha}(k+1)+\eta_{\alpha}(\varepsilon)
\ ,\label{epvarep}
\end{equation}
when $\varepsilon\leq\alpha^{1/(1-\alpha)}$. This follows from
Lemma \ref{fxy} and $\epsilon\leq{G}_{(k)}[\pq-\qq]\leq\varepsilon$ by
(\ref{kfcon}). Thus, we have arrived at a conclusion that
generalizes Fannes' inequality with respect to separate terms in
the expression for entropy.

\begin{Thm}\label{theo02}
For given $\alpha\in(0;2]$ and
$k\in\{1,\ldots,d\}$, if density operators $\bro$ and $\bva$
satisfy the condition
$\|{\bro}-{\bva}\|_{(k)}=\varepsilon\leq\alpha^{1/(1-\alpha)}$
then
\begin{equation}
\left|S_{\alpha}^{(k)}({\bro})-S_{\alpha}^{(k)}({\bva})\right|
\leq\varepsilon^{\alpha}\ln_{\alpha}(k+1)+
\eta_{\alpha}(\varepsilon)
\ . \label{inefan1}
\end{equation}
\end{Thm}

\section{Inequalities for the case $\alpha\in(2;+\infty)$}\label{alpinf}

The second method of estimate with restriction $\alpha>1$ is
presented in \cite{zhang}. Let $\pq$ and $\qq$ be $m$-dimensional
probability vectors. The $l_1$-distance (or Kolmogorov distance)
between them is defined as \cite{chuang}
\begin{equation}
D({\pq},{\qq}):=\frac{1}{2}\sum\nolimits_{i=1}^{m}|p_i-q_i|\equiv\frac{1}{2}\>G_{(m)}({\pq}-{\qq})
\ . \label{kolmpq}
\end{equation}
Using probabilistic coupling techniques, the following statement has been proved \cite{zhang}. For $\alpha>1$, there
holds
\begin{equation}
\left|H_{\alpha}^{(m)}(\pq)-H_{\alpha}^{(m)}(\qq)\right|\leq\delta^{\alpha}\ln_{\alpha}(m-1)+
H_{\alpha}(\delta,1-\delta)
\ , \label{reszhan}
\end{equation}
where $\delta=D({\pq},{\qq})$ and
$H_{\alpha}(\delta,1-\delta)\equiv\eta_{\alpha}(\delta)+\eta_{\alpha}(1-\delta)$
is the binary Tsallis entropy \cite{zhang}. The normalization
conditions $\sum_{i=1}^{m}p_i=\sum_{i=1}^{m}q_i=1$ are crucial for
use of probabilistic coupling. Nevertheless, we can utilize the
result (\ref{reszhan}) in obtaining estimates for partial sums of
Tsallis entropy.

\begin{Rem}\label{rgnd}
The right-hand side of (\ref{reszhan}) can be rewritten as $g(\delta,m-1)$, where we put
\begin{equation}
g(\delta,n):=\delta^{\alpha}\ln_{\alpha}(n)+
H_{\alpha}(\delta,1-\delta)=(1-\alpha)^{-1}\bigl(n^{1-\alpha}\delta^{\alpha}+(1-\delta)^{\alpha}-1\bigr)
\ . \label{gddef}
\end{equation}
By definition, we have $\eta_{\alpha}(\delta)\leq g(\delta,n)$. As
a function of $\delta$ at fixed $n$, the $g(\delta,n)$
monotonically increases in the range $0<\delta<{n/(n+1)}$. Indeed, for $\alpha>1$
the condition $\partial{g}/\partial\delta>0$ leads to
$(\delta/n)<(1-\delta)$.
\end{Rem}

\begin{Rem}\label{impab}
If $|B|\leq{C}$ and $A>0$ then $|A+B|\leq{C}$ implies $A\leq{C}+|B|$ inevitably.
\end{Rem}

\begin{Lem}
If
${\>}\epsilon\leq\min\left\{\alpha^{1/(1-\alpha)},(k+1)/(k+2)\right\}$
and $\alpha>1$ then in the domain $\mathfrak{D}_{\epsilon}$ it
holds that
\begin{equation}
|F({\xx},{\yy})|\leq g(\epsilon,k+1)+\eta_{\alpha}(\epsilon)
\ . \label{kpartz}
\end{equation}
\end{Lem}

{\it Proof.} With no loss of generality, the quantity
$F({\xx},{\yy})$ can be meant as positive. Consider the two cases.

(i) Components of $\xx$ and $\yy$ obey $u+\sum_{i=1}^{k}
x_i=\sum_{i=1}^{k} y_i=1-v$, where $u,v\geq0$. For $m=k+2$, we
build two $m$-dimensional probability vectors
$\pq=(x_1,\ldots,x_k,u,v)$ and $\qq=(y_1,\ldots,y_k,0,v)$.

(ii) Components of $\xx$ and $\yy$ obey $u+\sum_{i=1}^{k}
y_i=\sum_{i=1}^{k} x_i=1-v$, where $u,v\geq0$. For $m=k+2$, we
build two $m$-dimensional probability vectors
$\pq=(x_1,\ldots,x_k,0,v)$ and $\qq=(y_1,\ldots,y_k,u,v)$.

After cancellation of $\eta_{\alpha}(v)$, for both the cases we can write
\begin{equation}
H_{\alpha}^{(m)}(\pq)-H_{\alpha}^{(m)}(\qq)=F({\xx},{\yy})\pm\eta_{\alpha}(u)
\ , \label{combii}
\end{equation}
where the sign '$+$' is put in the case (i), the sign '$-$' is put
in the case (ii). In both the cases, we also have
$$
u=\left|\sum\nolimits_{i=1}^{k} x_i-\sum\nolimits_{i=1}^{k} y_i\right|
\leq\sum\nolimits_{i=1}^{k} |x_i-y_i|\leq
G_{(k)}({\xx}-{\yy})=\epsilon \ ,
$$
whence $2D({\pq},{\qq})=\sum_{i=1}^{k} |x_i-y_i|+u\leq2\epsilon$.
So we can take the result (\ref{reszhan}) with
$\delta\leq\epsilon$ and $m=k+2$. When $\epsilon\leq(k+1)/(k+2)$,
we have $g(\delta,k+1)\leq{g}(\epsilon,k+1)$ and
$$
|F({\xx},{\yy})+B|\leq g(\epsilon,k+1) \ .
$$
Here we use (\ref{combii}) and $B=\pm\eta_{\alpha}(u)$. If
$u\leq\epsilon\leq\min\left\{\alpha^{1/(1-\alpha)},(k+1)/(k+2)\right\}$ then
$|B|=\eta_{\alpha}(u)\leq\eta_{\alpha}(\epsilon)\leq{g}(\epsilon,k+1)$,
and the implication of Remark \ref{impab} completes the proof. $\square$

It must be stressed that the two ranges $\alpha\in(0;2]$ and
$\alpha\in(1;+\infty)$, in which the two estimates (\ref{kpart})
and (\ref{kpartz}) are respectively valid, have the joint interval
$(1;2]$. Here the bound (\ref{kpartz}) is weaker, and we shall use
it for $\alpha\in(2;+\infty)$ only. For small $\epsilon$, the
difference between these bounds is small though. Similar to the
inequality (\ref{inefan1}), the following result is immediately
obtained from (\ref{kpartz}).

\begin{Thm}\label{theo2in}
For given $\alpha\in(2;+\infty)$ and
$k\in\{1,\ldots,d\}$, if density operators $\bro$ and $\bva$
satisfy the condition
$\|{\bro}-{\bva}\|_{(k)}=\varepsilon\leq\min\left\{\alpha^{1/(1-\alpha)},(k+1)/(k+2)\right\}$
then
\begin{equation}
\left|S_{\alpha}^{(k)}({\bro})-S_{\alpha}^{(k)}({\bva})\right|
\leq\varepsilon^{\alpha}\ln_{\alpha}(k+1)+
\eta_{\alpha}(\varepsilon) +H_{\alpha}(\varepsilon,1-\varepsilon)
\ . \label{inefan2}
\end{equation}
\end{Thm}

\begin{Rem}
The inequalities (\ref{inefan1}) and
(\ref{inefan2}) are expressed in terms of the distances
$\|{\bro}-{\bva}\|_{(k)}$ that enjoy similar properties to the
standard trace distance \cite{rast091}. These measures are closely
related to Uhlmann's partial fidelities. In general, the fidelity
concept provides a natural extension of notion of transition
probability to mixed quantum states \cite{uhlmann76}. The
$k$-fidelity between ${\bro}$ and ${\bva}$ is defined as
\cite{uhlmann00}
$$
F_k(\bro,\bva):=\sum\nolimits_{j=k+1}^{d}
\sigma_j^{\downarrow}(\sqrt{\bro}\sqrt{\bva}) \ .
$$
It has been shown \cite{rast092} that
$\|{\bro}-{\bva}\|_{(k)}\leq2\left(1-F_k(\bro,\bva)\right)$. By
replacing $\varepsilon$ with
$\veps':=2\left(1-F_k(\bro,\bva)\right)$, we obtain the
reformulations of (\ref{inefan1}) and (\ref{inefan2}) in terms of
the partial fidelities under the corresponding conditions on
$\veps'$.
\end{Rem}

Theorems \ref{theo02} and \ref{theo2in} together fully cover the
range $\alpha\in(0;+\infty)$. Note that both the bounds explicitly
depend only on $k$ and not on the dimension of $\hh$. They
establish that all the partial sums of the Tsallis entropy are
continuous with respect to the trace norm. Indeed, for
$k=1,\ldots,d$, we have
$\|{\bro}-{\bva}\|_{(k)}\leq\|{\bro}-{\bva}\|_{(d)}\equiv{\rm{tr}}|{\bro}-{\bva}|$
by the definition (\ref{fannorm}). The inequalities
(\ref{inefan1}) and (\ref{inefan2}) are useful, since previous
extensions of Fannes' inequality dealt with the Tsallis entropy as
a whole. However, by itself a continuity of some functional does
not imply the same property for separate terms of the functional
(see, for example, the ratio of (\ref{dif1}) to (\ref{dif2})). As
it is shown above, the partial entropic sums are also continuous.
So the Tsallis entropy enjoys a continuity property that is more
subtle in character.

Finally, we discuss a stability of partial entropic sums.
In order for a measure to be experimentally robust, it is
necessary to put the stability criterion proposed by
Lesche \cite{lesche}. This criterion is posed as follows
\cite{abe02,zhang}. Let $\Phi$ be a state functional on the
probability simplex $\Delta_m$. We assume the set
$\{\Phi({\pq}):\>{\pq}\in\Delta_m\}$ of positive numbers to be
bounded above by the least bound $\Phi_M$. The
stability means that for every $\delta>0$, there exists
$\varepsilon>0$ such that
$|\Phi({\pq})-\Phi({\qq})|\cdot\Phi_M^{-1}\leq\delta$ whenever
$G_{(m)}({\pq}-{\qq})\leq\varepsilon$. Since $G_{(k)}({\pq}-{\qq})\leq G_{(m)}({\pq}-{\qq})$,
we have
\begin{equation}
\left|H_{\alpha}^{(k)}({\pq})-H_{\alpha}^{(k)}({\qq})\right|\leq
g(\varepsilon,k+1)+\eta_{\alpha}(\varepsilon)
\label{stab}
\end{equation}
provided that $G_{(m)}({\pq}-{\qq})=\varepsilon\leq\veps_0$, where
$\veps_0=\min\left\{\alpha^{1/(1-\alpha)},(k+1)/(k+2)\right\}$.
The relations (\ref{stab}) and (\ref{maxpts}) do enjoy the
stability property for partial entropic sums. Indeed, to each
$\xi\in(0;\veps_0)$ we assign
$$
\delta(\xi)=\bigl(g(\xi,k+1)+\eta_{\alpha}(\xi)\bigr)\ln_{\alpha}(k)^{-1}
\ , 
$$
so that the inequality
\begin{equation}
\left|H_{\alpha}^{(k)}({\pq})-H_{\alpha}^{(k)}({\qq})\right|
\left(\max{H_{\alpha}^{(k)}}\right)^{-1}\leq\delta(\xi)
\label{stabxi}
\end{equation}
is herewith stated for all $G_{(m)}({\pq}-{\qq})\leq\xi$. The
function $\delta(\xi)$ vanishes at the point $\xi=0$ and
monotonically increases with $\xi$ for $\xi\in(0;\veps_0)$ (see
Remarks \ref{econc} and \ref{rgnd}). So, one defines some
one-to-one correspondence between the intervals $[0;\veps_0]$ and
$[0;\delta(\veps_0)]$. Therefore, for each
$\delta\in(0;\delta(\veps_0)]$ we have $\xi(\delta)>0$ such that
the stability condition (\ref{stabxi}) does hold whenever
$G_{(m)}({\pq}-{\qq})\leq\xi(\delta)$. This is a particular
example to the fact that Lesche's stability property is formally
equivalent to uniform continuity \cite{abe02,zhang}.

\acknowledgments 
This work was supported in a part by the Ministry of Education and
Science of the Russian Federation under grants no. 2.2.1.1/1483,
2.1.1/1539. The present author is grateful to anonymous referees
for useful remarks.

\end{document}